# A CASE STUDY ON NUTEK INDIA LIMITED, REGARDING DEEP FALL IN SHARE PRICE

*Gurjeet Singh*,

Research Scholar, Computer Science,
Research Department,
Jaggannath University, Rajasthan, India

*Dr. Pankaj Nagar*,

Asstt. Professor, Department of Statistics,
University of Rajasthan, Jaipur
India

**ABSTRACT**

Manipulating the security price is an act of artificially inflating or deflating the price of a security. Generally, manipulation is defined as a series of transactions designed to raise or lower a price of a security or to give the appearance of trading for the purpose of inducing others to buy or sell. In essence, a manipulation is intentional interference with the free forces of supply and demand. In this paper we have tried to study the reasons behind drastic fall in share price of Nutek India Limited.

## 1. INTRODUCTION:

NTIL is a telecom infrastructure services company providing rollout solutions for both fixed and wireless telecom networks. NTIL provides expertise in turnkey site build, active equipment implementations, operations & maintenance and technical support services. The company's core expertise lies in the breadth of services it offers in the telecom infrastructure space. It offers services to telecommunication equipment manufacturers, telecom operators as well as third party infrastructure leasing companies in installing and maintaining telecom network equipment & Infrastructure. NTIL is also involved in creation of In-building networks for the Wireless and Data Applications. Its client list constitutes all the prominent players in the telecom industry that includes Third Party Infrastructure Leasing Companies (like Indus Towers, Quippo, WTTIL), Telecom operators (like Airtel, Vodafone, Idia, Reliance Communications, Aircel) and Telecom Equipment Manufacturers (like Ericsson, Nokia Siemens Network, Huawei, ZTE, Motorola)[16].

## 2. FACTS:

The share was listed at 192 rupees with FV 10 on 27[th] August 2008. In the mean time, splitted the FV to 5.00 (on 23[rd] December, 2009). The price of share fallen drastically from IPO listing price, Rs 192 to all time low of Rs 0.65 in BSE[12] and Rs 0.70 in NSE[13] on 25[th] Nov 2011. The book value of the share is Rs. 32.

**TABLE 1: MONTH WISE PRICE VARIATION IN NITL STOCK**

| Month | Highest Close Price | Month | Highest Close Price | Month | Highest Close Price |
|---|---|---|---|---|---|
| August-08 | 199.3 | November-09 | 96.15 | February-11 | 14.6 |
| September-08 | **207.6**** | December-09 | 91 | March-11 | 12.69 |
| October-08 | 80.05 | January-10 | 45.75 | April-11 | 12.72 |
| November-08 | 61.05 | February-10 | 35.95 | May-11 | 9.23 |
| December-08 | 55.05 | March-10 | 33.2 | June-11 | 5.98 |
| January-09 | 49.9 | April-10 | 42.75 | July-11 | 6.32 |
| February-09 | 41.5 | May-10 | 39.1 | August-11 | 4.31 |
| March-09 | 33.1 | June-10 | 34.85 | September-11 | 3.14 |
| April-09 | 47.95 | July-10 | 36.15 | October-11 | 3.28 |
| May-09 | 57.9 | August-10 | 38.85 | November-11 | 1.35 |
| June-09 | 59.55 | September-10 | 46.8 | December-11 | 1.48 |
| July-09 | 63.1 | October-10 | 60.35 | January-12 | 1.49 |
| August-09 | 63.05 | November-10 | 38.3 | February-12 | **1.12**** |
| September-09 | 107.1 | December-10 | 23.9 | | |
| October-09 | 103.4 | January-11 | 17.3 | | |

*: Lowest of 43 months   **: Highest of 43 months

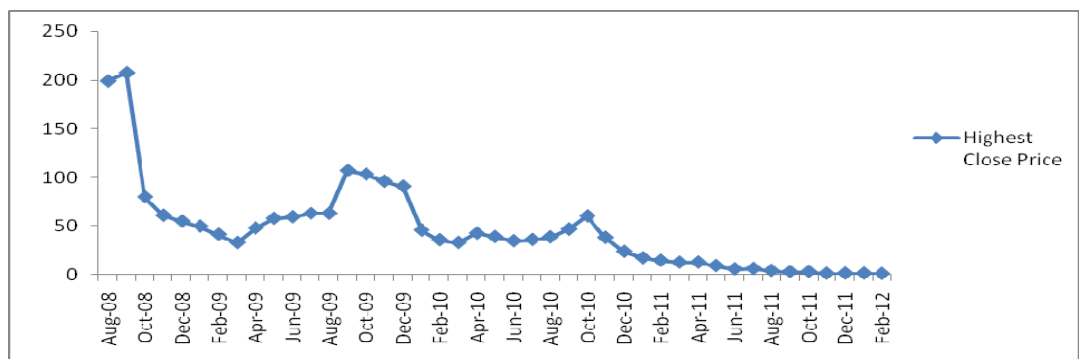

**Figure 1: Highest closing price of a month**

## 3. FACTORS:

Following are some factors which could be the predictors to detect the factual reason of drastic downfall in the stock price.

**(i) Price Manipulation by Operators:**

The bulk/block deals data is given below (Table 2) from NSE[13] since beginning of IPO and for the past one year from BSE[12]. Upon carefully analyzing the data, it is evident that following traders/individuals might have deliberately attempted to manipulate the share price while trading.

**TABLE 2: BULK/BLOCK DEALS OF IPO THROUGH DIFFERENT INVESTMENT ORGANIZATIONS**

| S.No. | Name of Trading Entity | Qty of Shares | | | Average Price (Rs.) of | | |
|---|---|---|---|---|---|---|---|
| | | Bought | Sold | Still Holding | Bought | Sold | Still Holding |
| 1 | ALFA FISCAL SERVICES PVT LTD (Trading Period-10th Nov. 2010 to 28th Oct., 2011). (Trading Avg. Price-Rs. 32.49 to 1.53) | 1,24,37,467 | 1,07,06,269 | 17,31,200 | 9.12 | 9.98 | 3.79 |
| 2 | DAVE CHETAN LAXMIKANT (Trading Period- 13th Oct 2011 to 28th Oct., 2011).(Trading Avg. Price-Rs. 3.76 to 1.5) | 1,09,85,164 | 92,89,511 | 16,95,653 | 2.39 | 2.2 | 3.45 |
| 3 | OVERALL FINANCIAL CONSULTANTS PRIVATE LIMITED (Trading Period-7th September, 2011 to 26th Oct., 2011).(Trading Avg. Price-Rs. 3.31 to 1.57) | 88,76,105 | 78,73,939 | 10,02,166 | 2.85 | 2.51 | 5.56 |
| 4 | TRANS FINANCIAL RESOURCES LIMITED* (Trading Period-28th September, 2010 to 16th March, 2011).(Trading Avg. Price-Rs. 61.75 to 10.95) | 3,11,99,132 | 3,12,82,848 | | 33 | 32 | Information is not available |
| 5 | PLANET INVESTMENTS & FINANCE PVT LIMITED (Trading Period-4th April, 2011 to 7th June, 2011).(Trading Avg. Price-Rs. 12.5 to 4.83) | 1,28,47,418 | 1,00,93,771 | 27,53,647 | 8.92 | 8.86 | 9.12 |
| 6 | VANRAJSINGH KAHOR (Trading Period-26th July, 2011 to 14th September, 2011). (Trading Avg. Price-Rs. 4.21 to 3.05) | 78,98,609 | 48,79,736 | 30,18,873 | 3.8 | 3.40 | 4.44 |
| 7 | SYNDICATE NIRMAN PVT.LTD (Trading Period-8th July, 2011 to 6th September, 2011).(Trading Avg. Price-Rs. 5.66 to 3.10) | 45,27,913 | 42,07,795 | 3,20,118 | 4.59 | 4.18 | 10.01 |
| 8 | KAPIL VIJAY AGRAWAL (Trading Period-9th Nov., 2010 to 24th Nov., 2010).(Trading Avg. Price-Rs. 36.23 to 23.89) | 65,48,810 | 62,48,810 | 3,00,000 | 30.81 | 30.60 | 35.12 |
| 9 | EPOCH SYNTHETIC PVT LTD (Trading Period-7th Nov., 2010 to 8th April, 2011).(Trading Avg. Price-Rs. 61.99 to 10.95) | 18,28,500 | 5,94,500 | 12,34,000 | 29.30 | 51.93 | 18.40 |
| 10 | KOYAL FINVEST PVT LTD (Trading Period-14th June, 2011 to 8th July, 2011).(Trading Avg. Price-Rs. 5.65 to 5.07) | 44,47,161 | 39,02,080 | 3,00,000 | 5.38 | 5.30 | 5.98 |

**\*:** More number of shares are sold than bought. May be due to discrepancies in input data. NSE data is from IPO date whereas BSE data is only of last one year.

**(ii) GDR Dumping:**

In the process total share capital increased by 12 cr and money close to Rs 316 Cr was raised by the company. During first GDR proceed, GDRs seem to have been issued at discounted prices in comparison to Indian market price, whereas the second GDR proceed looks to have been issued on par with the then market price.

**TABLE 3: TOTAL GDR ISSUED FOR FOREIGN INVESTMENT IN THE STOCK**

| Date | GDR Issued | Rate | Remarks |
|---|---|---|---|
| 5th Aug., 2010 | 40 Lac | USD 7.25 per GDR | This means addition of 4 cr shares to the share capital and money raised through this proceed is USD 29 Millions (which is equal to roughly Rs 125 Cr at 1 USD = Rs 43). The scrip was trading around Rs 35 per share in Indian market at that time. |
| 14th Dec., 2010 | Additional 80 Lac | USD 5.55 per GDR | This means addition of 8 cr shares to the share capital and money raised through this proceed is USD 44.4 Millions (which is equal to roughly Rs 191 Cr at 1 USD = Rs 43). The scrip was trading around Rs 24 per share in Indian market at that time. |

Money raised through GDRs at an average price close to Rs 30 per share. Total GDRs raised were 12 Cr share equivalents. Then price started falling down and ultimately saw its bottom at 65 paise on 25th Nov 2011.Price was manipulated downward while GDRs were being dumped in to the market. On 2nd Nov 2011, only 16.76% of GDRs were there with the custodians. Shares available at dirt cheap prices were accumulated and being converted to GDRs now. From 2nd Nov till date, GDRs with custodians has increased from close to 17% to 61%. GDR data as updated on 15th March 2012, it has increased from 52.07% to 61% of total share capital. In terms of total GDRs, it is 78.58%. Current share equivalent of GDRs is 9,429,951 out of 12 Cr total GDRs raised and against total share capital of 15,45,18,600. Without any surprise, the latest GDR holding is 61%, The detail are shown in Table 4 below:

**TABLE 4: GDR RISE**

| S.No. | Date | GDR Value |
|---|---|---|
| 1 | 2nd Nov 2011 | 16.78% |
| 2 | 8th Nov 2011 | 23.46% |
| 3 | 31st Dec 2011 | 27.97% |
| 4 | 12th Jan 2012 | 45.48% |
| 5 | 16th Feb 2012 | 52.07% |
| 6 | 15th March 2012 | 61 % |

According to **Table 3 & 4**, one can predict that GDRs were dumped at higher levels and bought back at dirt cheap prices. Now the question arises: Why does a GDR share holder sell at price less than the buying price and why does he again try to buy at much lower levels?

**(iii) Management is not trustworthy to investors:**

Management announced the Board Meeting to occur in August 2011, to consider the Bonus issue, twice they postponed the meeting and the reason they are giving is non availability of Directors[16]. If ever they had given some genuine reason, the investors could have understood. The said reason is as if they are fooling the investors.

- When the investors could see fraudulent trading entities resorting to deliberate price manipulation of the scrip by means of bulk deals, why management observed silence?
- Price was hitting almost continuous lower circuits in the past couple of months till it reached all-time low of Rs 0.65 in BSE on 25th Nov 2011. Management knows the names of the trading entities possibly involved in price manipulation, upon analyzing trading data from NSE & BSE for more than an year. What did the management do after knowing the names of those trading entities?
- The investors expected from the management to at least approach the regulator in ensuring that such entities be banned from further trading. Hence, management's claim that they value investors/shareholders interest is seen as not convincing.

- Management`s claim that there is no evidence on insider trading needs to be further confirmed by disclosing holdings of promoters, management, their friends and relatives (both in India and abroad) since capital is raised through GDR proceeds, to make the investors fully convinced.
- Investors strongly feel that, had the management come public on business channels and provided future performance guidance immediately after price fell below its face value, the situation would have not worsened as is the case now.
- Management`s claim that telecom industry turmoil as the possible reason is not convincing because other peers performance is not affected equally adversely.
- Though the claim of diversification by management looks good, why it took so long to provide critical information such as this earlier, when the price fell below its face value.
- Investors are also seeking answers to management`s decision to consider bonus issue, and then decision on disapproving the same. Investors were shell shocked to know that the board meeting got postponed twice and later got cancelled due to non-availability of its directors. This shows how committed the management is in investor interests.

(iv) **SEBI's Investigation:**

There are lot of complains[15] of investors against Nutek India Limited regarding price manipulation and deep fall in share price. SEBI investigation, is going on. In future we shall work on the issue and try to investigate on conceptual/Financial/ economic point of view and prepare a statistical report to examine whether there is some fraud/ manhandling in the stock is made.

Some research work and reports in this direction, have been mentioned in references [1]-[11] below.

**4. Conclusion:**

The principle goal of the present research is to offer the detection of price manipulation in a security, in this context we have studied and analyzed the data of Bulk Deals and GDR data of Nutek India Limited and we have seen how the share price fallen from its IPO listing price Rs 192 to Rs 0.65 due to price manipulation by operators, how GDR were dumped at higher levels and bought back at dirt cheap prices. Management made false statements regarding bonus issue and never bothered about deep fall in share price since IPO, hence Management is not trustworthy to investors. SEBI investigation is going on, yet the outcome of investigation has not come.

**REFERENCES:**

[1] Mary-Jo Kranacher, Richard Riley, Joseph T. Wells Forensic Accounting and Fraud Examination, John Wiley and Sons, 08-Jun-2010
[2] Joseph T. Wells ,Principles of fraud examination John Wiley, 2005
[3] Joseph T. Wells, Fraud examination, investigative and audit procedures, Quorum Books, 1992
[4] Jack C. Robertson, Timothy J. Louwers, Auditing, Irwin/McGraw-Hill, 1999 .
[5] David Coderre, Computer Aided Fraud Prevention and Detection, A Step by Step Guide, John Wiley and Sons,2009
[6] Albert S. Kyle and S. Viswanathan, How to Define Illegal Price Manipulation, January 14, 2008, www.duke.edu/.../PP-Article-Kyle_Vish_Manipulation_20080100_f.pdf
[7] Franklin Allen, Lubomir Litov, and Jianping Mei, Large Investors, Price Manipulation, and Limits to Arbitrage: An Anatomy of Market Corners, Draft of 1/14/2006, fic.wharton.upenn.edu/fic/papers/06/0602.pdf.
[8] Zakia Ferdousi† and Akira Maeda‡, Unsupervised Fraud Detection in Time Series data, www.inf.ufrgs.br/~alvares/CMP259DCBD/outlier_detection_PGA.pdf
[9] Bolton, R. J. and Hand. D.J. "Unsupervised Profiling Methods for Fraud detection,"Credit Scoring and Credit Control VII, Edinburgh, UK, 5-7 Sept (2001).
[10] Rajesh K. Aggarwal & Guojun Wu. Stock Market Manipulation — Theory and Evidence, Working paper, Univ. of Michigan. March 11,( 2003).
[11] T. Jeffery Wilks and Mark F. Zimbelman, Using Game Theory and Strategic Reasoning Concepts to Prevent and Detect Fraud, Brigham Young University, Accounting Horizons, Vol. 18, No. 3, September 2004.

**SOURCE OF DATA:**
[1] Bombay Stock Exchange's Website www.bse.com
[2] National Stock Exchange's Website www.nse.com
[3] Bank of New York Mellon Website www.adrbnymellon.com
[4] SEBI's Investors Complain Website www.scores.gov.in
[5] Nutek India Limited Website www.nutek. in

----